%% file: main.tex
\documentclass{iitgthesis}

\usepackage{epsfig}
\usepackage{stackrel}
\usepackage{epsf}
\usepackage{amssymb}
\usepackage{amsthm}
\usepackage{graphicx}
\usepackage[english]{babel}
\usepackage{mathrsfs}
\usepackage{fancyhdr}
\usepackage{verbatim}
\usepackage{amsmath,amssymb}

\usepackage{fixltx2e} 
\usepackage{algorithm}
\usepackage{algorithmic}
\usepackage{ragged2e}
\usepackage{comment}
\usepackage[toc,page]{appendix}
\usepackage[hidelinks]{hyperref}
\newcommand{\argmin}{\arg\!\min}
\newtheorem{defn}{Definition}

\usepackage{cite}
\usepackage{multirow,tabularx}
\usepackage{ifthen}
\usepackage{times}
\DeclareMathAlphabet{\mathpzc}{OT1}{pzc}{m}{it}

\usepackage{amssymb}
\usepackage{stmaryrd}
\usepackage{amsmath}
\usepackage{mathrsfs}

\begin{document}

\include{prelude}

\include{introduction}        
\include{Relatedwork}
\include{Implementationsimulation}
\include{Selective}

\include{Proposedscheme}

\include{Tokenconstruction}

\include{Security}
\include{Discussion}
\include{Simulations}
\include{Conclusion}
\appendix
\include{Appendix}






\begin{acknowledgments}
We would like to thank our guide Professor Kannan Karthik for his valuable inputs which were instrumental in formulating our proposal. We would also like to thank our friends for their encouragement and support. We would like to reserve special acknowledgement for Rahul Nallamothu and Mridul Krishna for their useful ideas during the problem formulation. We also wish to acknowledge our parents who motivated and supported us throughout the project.
\end{acknowledgments}

\end{document}

%% file: prelude.tex

\clearpage\pagenumbering{roman}  

\title{Centre driven Controlled Evolution of Wireless Virtual
Networks based on Broadcast Tokens}
\author{
Atishay Jain , Vignesh Babu
}
\date{April 2014}

\rollnum{10010271, 10010261}

\iitgdegree{Bachelor of Technology}

\thesis

\department{DEPARTMENT OF ELECTRONICS \& ELECTRICAL ENGINEERING}

\setguide{Dr. Kannan Karthik}

\maketitle

\makeapproval



\begin{abstract}

In a wireless sensor network, the virtual connectivity between nodes is a function of the keys shared between various nodes. Pre-embedding these key configurations in the nodes would make the network inflexible. On the other hand, permitting subsets of nodes to engage in a common key synthesis phase to create secure distributed connections amongst themselves, would decouple and conceal the information flow from the controlling centre. An intermediate solution is the notion of a centre driven key generation process through broadcast tokens, designed to extract different keys in different nodes based on some prior information stored at the nodes. As more tokens arrive, the virtual connectivity of the nodes are altered and the network evolves. This evolution can be distributed and can be controlled to converge to a certain specific connectivity profile. In this paper we present a framework and an algorithm which controls the simultaneous and distributed key release in different nodes, resulting in the creation of parallel virtual multicast groups. The design of the node shares and the supporting broadcast tokens have been discussed in conjunction with the process of balancing the spans of individual groups with spans of several coexistent multicast groups.

\end{abstract}

\tableofcontents
\listoffigures




\cleardoublepage\pagenumbering{arabic} 

%% file: introduction.tex
\chapter{Introduction}
\label{chap:intro}
In collaborative information processing applications, a large computational task is delegated amongst several wireless nodes, by a centre. The main task is split into several sub-tasks and each subtask is assigned to a different group of nodes. Since the computation of each subtask entails a sharing of resources and may require further modularization, the nodes can form virtual groups for exchanging information and also for delegating roles amongst themselves. The information flow across virtual groups, group expansions, group migrations, group mergers, group dissolutions etc. can either be completely distributed or can be triggered by the centre. However when the nodes carry confidential information (related to a large product design) and the overall computational task is also of a sensitive nature, the centre cannot afford to allow the nodes to form their own sub-groups to perform this distributed processing. This would make it difficult for the centre to monitor the flow and also perform intermediate checks on the information fragments exchanged amongst the nodes. Hence, a centre directed group formation and information exchange is recommended. \\ \\
Secure virtual multicast connections either between subsets of nodes and/or dedicated unicast links between each of the nodes and the centre \textit{C}, are required to preserve confidentiality of messages exchanged between the nodes within a specific virtual group or between the nodes and the centre. Any secure connection, requires the sharing of an encryption key, which can be pre-distributed by the centre at the time of forming the network and registering new nodes. Alternatively the centre may facilitate the generation of keys in a distributed fashion between several clusters of nodes. The strength of this key establishment phase or key distribution phase is important in determining the overall security of the network. Several key distribution mechanisms for Wireless Sensor Networks have been studied in the past.\\ \\
In pre-distributed shared key mechanisms \cite{Lee05}\cite{Lee052}, the centre pre-assigns a set of keys to each node before they are deployed. The key pre- distribution is such that any two neighbouring nodes share a single common key or a group of common keys with a certain probability.  In Random pair-wise key schemes \cite{Liu05}, a random set of node IDs and corresponding link keys are stored in each node prior to deployment. The nodes broadcast these IDs to advertise the presence of common keys and the connectivity graph is created. An important drawback of these approaches is the degree of uncertainty in connectivity which could lead to disconnected networks. Location based Pair wise key distribution schemes \cite{Liu03} assume that the network topology is known beforehand and the keys stored in each node depend on its neighbours. Hence any two nodes are guaranteed to share a common key. However, these schemes assume that the topology is known prior to deployment, which is rarely the case in practical scenarios. The main problem with these approaches is that physical capture of a node could lead to the loss of many keys and link associations.\\ \\
Key matrix based schemes described in \cite{Blom85} define a key matrix which stores all the link keys. Each node is given some public and private information derived from the matrix. Neighbouring nodes exchange their public information which is then combined with their respective private information to generate the link key. In Polynomial based key distribution approaches \cite{Blundo93}, each node stores a partially evaluated symmetric polynomial which is evaluated using the IDs of the neighbouring nodes to establish link keys. Above dynamic key distribution methods, avert problems associated with node-capturing in static pre-distribution schemes. However these approaches are power hungry, computationally intensive and require costly vector multiplications. \\ \\
Broadcast encryption schemes allow transmission of secrets to a privileged set through a broadcast channel \cite{Fiat94}. Broadcast encryption techniques are used in schemes based on combinatorial constructions, one-time revocation schemes based on secret sharing techniques, and tree-based constructions which have been proposed in \cite{Kogan03}. Broadcast encryption schemes define a parameter $k$ which specifies that at least $k$ users who do not belong to the privileged set have to collude in order to decrypt the message. Hence they are generally only $k$-resilient. Moreover they incur high communication cost to broadcast messages, large memory requirement to store keys in each node and limited group association structures. Further, in all of these schemes, the establishment of group keys for an arbitrary privileged set of size $m$, would generally require the exchange of at the most $m$ messages resulting in wastage of the available bandwidth. If $N_G$ simultaneous multicast groups must be created, the number of messages required would be anywhere between $N_G$ and $m\times N_G$. Tree based constructions used in Broadcast encryption, have been designed to counter multicast group dynamics and have been optimized for single multicast groups. Simultaneous creation of multiple virtual multicast groups would however require messages proportional to the number of groups. \\ \\
Another common problem with the above mentioned approaches is that the configuration of the network is static which means that the virtual connections within and between subsets of nodes in the network cannot change with time. However in a collaborative processing application, the topology of the network once deployed may have to be reconfigured to facilitate parallel and distributed computation and also to ensure that intermediate computational results are exchanged amongst different virtual groups. Hence, dynamic re-configurability of the virtual network is paramount. However, when the information handled by the nodes is of a sensitive nature, this re-configuration must be triggered by the centre, to ensure transparency in the information flow.\\ \\
Another approach, to address this issue of centre driven re-configurability~\cite{Karthik13}, considered an interaction between protected node shares of key blocks and broadcast tokens released by the centre. Here, each wireless node in the network was assigned a protected node-share of an encryption/decryption key set. Upon the release of specially designed tokens which are broadcasted by a centre, the fusion of these shares with the tokens would unlock a set of encryption keys. Common keys unlocked in different nodes can be used to form multicast group associations. The associations can then be easily changed by broadcasting new tokens. The scheme is computationally feasible, fully resilient and physical node capture only results in the loss of the keys which have been unlocked in the captured node. Thus a single broadcast token can result in the formation of several multicast groups, the choice of groups and their evolution can be pre-designed to satisfy a certain function. This pre-design retards the flexibility of the network.\\ \\
However, it did not specify a design to control the key release and most importantly alter it dynamically to suit a certain network computational goal. A solution which allows the centre to design tokens that unlock desired keys (controlled key release) in different nodes is imperative to achieve any desired configuration. In this report we propose such a solution based on the protocol described in \cite{Karthik13}. \\ \\
The thesis report is organised as follows: In chapter 2, the virtual reconfiguration scheme proposed in \cite{Karthik13} is discussed. Chapter 3 provides details of the software implementation of the reconfiguration protocol. In chapter 4, we discuss our initial approach to incorporate controlled key release.  In chapter 5, we discuss the framework and the proposed solution which improves up on the solution discussed in chapter 4 and achieves controlled and simultaneous key release. Algorithms which use the proposed solution for efficient Token construction to achieve simultaneous creation of multicast groups is discussed in Chapter 6. Issues pertaining to key leakage and token collusions are discussed in chapter 7. Comparison with tree broadcast encryption and extensions towards forward secrecy are explored in chapter 8. Finally, in chapter 9, some simulation results are presented.

%% file: Relatedwork.tex
\chapter{Virtual Network Reconfiguration}
\label{chap:VirtualNetworkReconfig}
A protocol based on key release mechanisms to reconfigure networks has been discussed in detail in \cite{Karthik13}. The entire protocol has been described here for the purpose of clarity which would be useful in understanding subsequent sections. The network is assumed to have a distribution center and a certain number of nodes. The scheme revolves around the idea that a set of encryption keys are locked within protected node-shares in each node of the network. When specially designed tokens are broadcasted by the center, the fusion of these shares with the tokens release a subset of the locked keys. With the arrival of every new token more keys would be released at each node. If the node-shares stored in different nodes are dissimilar, it would imply that different sets of keys could be released by the same token in different nodes. The unlocked keys would then determine the configuration of the network. If a key is common to a set of nodes it would necessarily imply a multicast connection between them in the sense that this key can be used to transmit messages securely within the group formed. With the broadcast of each new token, the configuration of the network changes dynamically, as the associations between different nodes change owing to the release of new keys in each node.

\section{Description of the model}

The model considers a distribution center \textit{C} and $n$ number of nodes say Node $1$, Node $2$, Node $3$, $\dots$ , Node $n$. We assume that all the nodes including the center are in the transmission range of any other node in the network.
The center generates a set of $v$ keys say $K\textsubscript{1}$, $K\textsubscript{2}$, $K\textsubscript{3}$,$\dots,\; K\textsubscript{v}$ and a set of shares $N\textsubscript{1}$, $N\textsubscript{2}$, $N\textsubscript{3}$, $\dots,\; N\textsubscript{n}$ for each node in the network. This means that the $i\textsuperscript{th}$ node is given the node-share $N\textsubscript{i}$ prior to deployment. The center also generates and broadcasts specially designed tokens $T\textsubscript{k}\textsuperscript{'}s$. The node-shares $N\textsubscript{i}\textsuperscript{'}s$ have complete information required to extract all the $v$ keys using the broadcast tokens $T\textsubscript{k}\textsuperscript{'}s$. The node-shares $N\textsubscript{i}\textsuperscript{'}s$ are different for different nodes and every broadcast token $T\textsubscript{k}$ upon fusion with the node-shares, unlocks different subset of keys in different nodes of the network. This results in the formation of different clusters of nodes each with common shared keys which can be further used for secure multicast communications.

\section{Generation of node-shares and broadcast tokens}

The distribution centre generates these node-shares and tokens using a non-perfect secret sharing scheme called the MIX-SPLIT \cite{Karthik08}.
It assumes the keys to be uniform length $L\textsubscript{p}$-bit random strings. A block $X$ of length ($L\textsubscript{p} \times v$ ) bits is computed as follows. The keys $K\textsubscript{1}$, $K\textsubscript{2}$, $K\textsubscript{3}$, $\dots,\; K\textsubscript{v}$ are first concatenated into a string and interleaved without changing the order of the bits, to form a block $X$. The partition of a key is defined as the set of bit positions in $X$ that are filled with the bits of that particular key in the same order. Since there are $v$ keys, there would be $v$ disjoint partitions which are designated as $P\textsubscript{1},\; P\textsubscript{2},\; P\textsubscript{3}, \dots,\; P\textsubscript{v}$ each of length $L\textsubscript{p}$. These partitions are also referred to as hidden partitions as they are unknown to the nodes. Another block $Y$ is defined as the bit complement of $X$ (bit wise not of $X$).\\ \\
Macro-mixing of fragments of $X$ and $Y$  is done to produce $r$ preliminary shares say $PS\textsubscript{1},\: PS\textsubscript{2},\\ \dots \:PS\textsubscript{r}$. Subsets of fixed size from these preliminary shares in turn form different node share matrices $N\textsubscript{i}’s$. Each of these preliminary shares can be written in the following form:
\begin{equation}
PS\textsubscript{i} = (PS\textsubscript{i1} || PS\textsubscript{i2} || PS\textsubscript{i3} ||\dots PS\textsubscript{iv}) \label{eqn:Si}
\end{equation}
where the sub-sequence PS\textsubscript{ij} is derived according to a pre-defined codebook $C$ defined as
\[
C =
\left({\begin{array}{cccc}
c_{11} & c_{12} & \dots & c_{1v} \\
c_{21} & c_{22} & \dots & c_{2v} \\
\dots & \dots & \dots & \dots\\
c_{v1} & c_{v2} & c_{v3} & c_{vv} 
\end{array} } \right)
\]
$PS\textsubscript{ij}$ is obtained as follows\\
\begin{equation*}
PS\textsubscript{ij} = X(P\textsubscript{j})\;\;\;if\;\;c\textsubscript{ij} = 1\\
\end{equation*}
\begin{equation*}
PS\textsubscript{ij} = Y(P\textsubscript{j})\;\;\;if\;\;c\textsubscript{ij} = 0
\end{equation*}
where $c\textsubscript{ij}$ is an entry of the code book. In general the code book can be partitioned as
\[
C =
\left({\begin{array}{c}
N \\
\hline
T
\end{array} } \right)
\]
where $T$ is a $t \times v$ matrix whose rows are $t$ code words which are used to construct the broadcast tokens and $N$ is a $(r\;–-\;t) \times v$ matrix also called as the node-share matrix, whose rows are code words which are used to construct the preliminary shares $PS\textsubscript{i}$. The preliminary node shares and broadcast tokens are built from their respective code words by the above mentioned method. Subsets of these preliminary shares in turn form node-shares which are distributed to all the nodes Node 1, Node 2, Node 3, \dots Node $n$. \\ \\
The operator $\parallel$ stands for the concatenation and mixing operation. The values in the bit positions specified by the partition $P\textsubscript{j}$ for a particular sub-sequence $PS\textsubscript{ij}$, are chosen from $X$ or $Y$ depending on the value of $c\textsubscript{i,j}$. In other words, the values of the bit positions in $PS\textsubscript{i}$ which are specified by $P\textsubscript{j}$ will either be equal to the bits of $K\textsubscript{j}$ or $K\textsubscript{j}\textsuperscript{c}$ depending on the value of $c\textsubscript{ij}$. This procedure is then repeated $v$ times for each sub-sequence $PS\textsubscript{ij}$ to form the preliminary share $PS\textsubscript{i}$. The figures given below illustrate the mixing process for a particular 3 key system where each key is 4 bits long.

\begin{figure*}[tbh]
\centering
\includegraphics[totalheight = 0.25\textheight, width=0.6\textwidth]{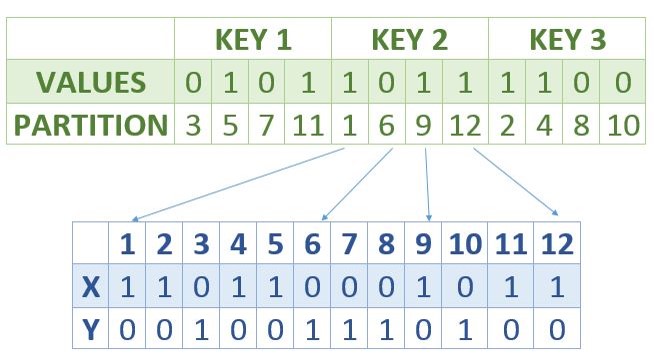}                            
\label{fig:figure1}
\caption{An example of partition assignment in a 3 key system}
\end{figure*}

\begin{figure*}
\centering
\includegraphics[totalheight = 0.25\textheight, width=0.6\textwidth]{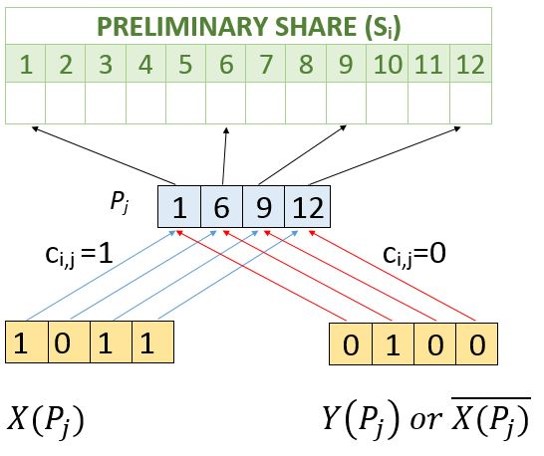}
\label{fig:figure2}
\caption{Illustration of the key mixing process}
\end{figure*}

\section{Rules for the design of code books}

Since information regarding all the $v$ keys are contained in each of the preliminary shares, it is possible to extract a subset of these keys by stacking selective shares one above the other \cite{Karthik12}. The rules for the design of a code book to enforce conditional visibility and invisibility of partitions (for unlocking a subset of keys) are as follows:\\ \\
{\bf Rule 1:} 
Complementary and repetitive columns lead to inseparable partitions i.e, if a set of code words which when stacked over each other form a matrix whose columns are either repetitive or complementary, then the preliminary shares or tokens generated from these code words do not reveal any hidden partitions.\\ \\
{\bf Rule 2:} 
Row-sampling of a complementary pattern is complementary. In other words, if a subset of code words are chosen from a set of code words satisfying the complementary column property, then these subset of code words would also satisfy the complementary column property. \\ \\
{\bf Rule 3:} 
Single shares are always mixed i.e, no partitions are revealed by one share alone. Multiple shares/tokens have to be stacked to reveal partitions. \\ \\
{\bf Rule 4:} 
At least one partition becomes visible if a column is distinct. Hence, if the code-words which are stacked, form a matrix with a unique column, then that corresponding partition and hence that key, can be unlocked by stacking the actual shares/tokens generated by these code words and searching for that unique column pattern. \\ \\
The node shares that are distributed to nodes in the network, must not reveal any keys on their own. Hence, the corresponding node share code words should have complementary or repetitive columns (Rule 1). The keys should only be released upon fusion with the broad casted tokens. This means that the stacking of node share code words plus the token code words should form a matrix with one or more unique columns (Rule 4). An additional security requirement that cannot be ignored is that the Tokens themselves should also satisfy Rule 1 to ensure that stacking any subset of Tokens does not reveal hidden partitions to an eavesdropper. 

\section{Revealing hidden partitions}
When the stacked code words reveal unique columns, these unique columns can be converted into distinct bit patterns. For instance, if the unique column formed by stacking 3 code words is the binary number $[1\: 0\: 1]\textsuperscript{T}$, then it represents the bit pattern $[b\: \bar{b} \: b]\textsuperscript{T}$. In order to reveal the corresponding hidden partition in the actual shares (shares generated by the three code words), this bit pattern is searched on these shares. In other words, the actual shares are traversed column by column and the set of indexes of every column which represents this bit pattern, is the hidden partition \cite{Karthik12}\cite{Karthik08}. In this example, the hidden partition corresponds to the column indexes in the shares where the following condition is satisfied: 
\[
Share1(i)\: =\: Share2(i)\textsuperscript{c}\:=\: Share3(i)
\]
All the bit positions $i$ which satisfy this condition form the hidden partition.

\section{Illustration}
Consider a case with a center C and 3 nodes in the network. Let us consider a node-share matrix to be as follows:\\
\[
N =
\left({\begin{array}{cccccc}
1 & 1 & 0 & 0 & 0 & 1 \\
0 & 1 & 1 & 1 & 0 & 0 \\
1 & 0 & 0 & 1 & 1 & 0 
\end{array} } \right)
\]
\\
It can be observed that the above matrix satisfies Rule 1 i.e, no columns are unique. They are either the same or complementary to each other. Using the above matrix let us construct the node share code word matrices $SH\textsubscript{1}$, $SH\textsubscript{2}$, $SH\textsubscript{3}$ for the 3 nodes as follows:\\ 
\[
SH\textsubscript{1} =
\left({\begin{array}{cccccc}
1 & 1 & 0 & 0 & 0 & 1 \\
0 & 1 & 1 & 1 & 0 & 0 \\
\end{array} } \right)
\]
\[
SH\textsubscript{2} =
\left({\begin{array}{cccccc}
0 & 1 & 1 & 1 & 0 & 0 \\
1 & 0 & 0 & 1 & 1 & 0 \\
\end{array} } \right)
\]
\[
SH\textsubscript{3} =
\left({\begin{array}{cccccc}
1 & 1 & 0 & 0 & 0 & 1 \\
1 & 0 & 0 & 1 & 1 & 0 \\
\end{array} } \right)
\]
\\
It can be noted here that these matrices are obtained by considering code word subsets of size 2. It is again imperative that the broadcast tokens T\textsubscript{k}’s also satisfy Rule 1 as they are exposed to traitors not belonging to the network. Our aim is to release some of the keys upon the fusion of tokens with the node shares. Hence we consider a token generator matrix with repetitive columns (to satisfy Rule 1). All the token code words obey the following format.\\
\[
T =
\left({\begin{array}{cccccc}
z_{1} & z_{2} & z_{3} & z_{2} & z_{3} & z_{1} \\
\end{array} } \right)
\]
\\
The total number of arrangements of the bit-positions is $2\textsuperscript{3}\: =\: 8$, the total number of unique tokens satisfying Rule 1 is 8. Now this number also includes those token code words which are bit complimentary with respect to other tokens. These tokens will not change the stack relation and are redundant. Hence the number of useful tokens is  $2\textsuperscript{3}/ 2 \:=\:4$. Accordingly, the token generator matrix can be represented as follows:\\
\[
T =
\left({\begin{array}{cccccc}
1 & 1 & 0 & 1 & 0 & 1 \\
1 & 0 & 1 & 0 & 1 & 1 \\
1 & 1 & 1 & 1 & 1 & 1 \\ 
1 & 0 & 0 & 0 & 0 & 1 \\
\end{array} } \right)
\]
\\
where each row represents a broadcast token code word. The actual broadcast tokens are generated from these code words. When a Token for example $T\textsubscript{1}$ is broadcasted , at Node $1$ it will fuse with the node-share corresponding to Node $1$ which is $SH\textsubscript{1}$ and will give rise to the following three distinct stack equations:\\
\[
\left({\begin{array}{c}
b \\
b \\
b \\
\end{array} } \right),
\left({\begin{array}{c}
b \\
b \\
\bar{b} \\
\end{array} } \right), 
\left({\begin{array}{c}
b \\
\bar{b} \\
b \\
\end{array} } \right)
\]
\\ 
\[
T\textsubscript{1} + SH\textsubscript{1} =
\left({\begin{array}{cccccc}
1 & 1 & 0 & 1 & 0 & 1 \\
1 & 1 & 0 & 0 & 0 & 1 \\
0 & 1 & 1 & 1 & 0 & 0 \\
\end{array} } \right)
\]
Out of these three equations only one is unique which is 
\[
\left({\begin{array}{c}
b \\
\bar{b} \\
b \\
\end{array} } \right)
\]
This implies that the partition $P\textsubscript{4}$ will be made visible. To obtain key 4 corresponding to this partition we traverse through the stack of actual shares/tokens (generated using the MIX SPLIT method by substituting the code-word bits by the actual bits of X or Y) and check for the pattern given above \cite{Karthik08}. The indexes which satisfy this pattern will give us the desired bit locations. The values at these bit locations when concatenated will give us either key $K\textsubscript{4}$ or its compliment. Hence key $4$ is unlocked. Similarly, this method is followed in all the other nodes as well. The unlocked partitions in Node $2$ and Node $3$ are $6$ and $4$ respectively. The set of unlocked keys at each node increases with the arrival of new tokens. The table below shows the sequential unlocking process with the arrival of each token.
\begin{table}[tbh]
    \centering
    \caption{Cumulative list of keys unlocked at each node}
    \begin{tabular}{|c|c|c|c|}
    \hline
    {\bf Token no} & {\bf Node 1} & {\bf Node 2} & {\bf Node 3}\\
    \hline
    1 & K\textsubscript{4}  & K\textsubscript{6} & K\textsubscript{4}               \\
    \hline
    2 & K\textsubscript{3},K\textsubscript{4}  & K\textsubscript{3},K\textsubscript{6} & K\textsubscript{1},K\textsubscript{3},K\textsubscript{4}               \\
    \hline
    3 & K\textsubscript{2},K\textsubscript{3},K\textsubscript{4},K\textsubscript{5}  & K\textsubscript{3},K\textsubscript{4},K\textsubscript{6} & K\textsubscript{1},K\textsubscript{3},K\textsubscript{4}               \\
    \hline
    4 & K\textsubscript{2},K\textsubscript{3},K\textsubscript{4},K\textsubscript{5}  & K\textsubscript{3},K\textsubscript{4},K\textsubscript{5},K\textsubscript{6} & K\textsubscript{1},K\textsubscript{2},K\textsubscript{3},K\textsubscript{4}               \\
    \hline
    \end{tabular}
    
\end{table}
\\

%% file: Implementationsimulation.tex
\chapter{Implementation}
\label{chap: ImplementSimulate}

\section{Design environment and testing}
We implemented the protocol on the contiki-2.6 operating system. It has been simulated on the IEEE 802.15.4 based TMote Sky platform. The Cooja simulator that comes in built with the Contiki OS, was used for simulating the wireless sensor network (WSN). The programming was done in C language.
\section{Implementation specifics}
\begin{itemize}
\item \textit{Inbuilt code book}: The code book is stored in the distribution centre prior to the deployment of the network.
\item \textit{Initial Node share distribution}: In our implementation, after the network simulation is started, the centre computes the Node shares and unicasts them to the respective nodes. We assume that network is deployed only after the initial node shares have been distributed by the centre. Hence, the eaves dropper cannot obtain these node shares.
\item \textit{Token Broadcast}: The centre broadcasts the Tokens periodically. Hence, the configuration of the network will change at regular intervals.
\item \textit{Encryption/Decryption algorithm}: We implemented and tested the Tiny Encryption Algorithm (TEA) along with this protocol. This symmetric key encryption/decryption algorithm is optimum for small wireless device security. It can be used for actual encryption/decryption of messages that are exchanged between nodes once the reconfiguration protocol unlocks some common keys. Provisions have been made for the messages to be encrypted and sent in numbered fragments. Acknowledgements are sent back to the sender upon successful reception. Duplicate packets are also removed automatically.
\end{itemize}
\section{Simulations}
We simulated the protocol for a 6 key 4 node network. One of the nodes in the network was programmed to behave as the distribution centre. All the nodes were within the transmission range of each other. We designed a code book and verified the pattern of keys released with each new broadcast token. The protocol was observed to function properly and changes in the configuration changes with each Token were observed to be correct.\\
After a thorough investigation, we were able to identify the main strengths and weaknesses of the described scheme.
\section{Protocol analysis}
\textbf{Strengths}:\\
The main strengths of the protocol are:
\begin{itemize}
\item Enables virtual re-configurability of the network by using a center driven broadcast system. For example, it can be used in a WSN to re-distribute load adaptively within the network.
\item For a network with a large number of keys, the damage caused by capturing certain nodes can be reduced since only a small subset of keys will be compromised.
\item Secure additions of nodes to a multicast group without leakage of information.
\item Secure unicast connections with the center are also possible.
\end{itemize}
\textbf{Drawbacks}:
\begin{itemize}
\item \textit{Configuration Control} - There is no control over the keys which are unlocked by the tokens in each node and hence we have no control over the configuration change.
\item \textit{Possible Configurations} - The number of tokens that can be broadcasted, are limited by the number of keys. For example only 4 tokens are possible for 6 keys. Since we do not have control over the release of keys and tokens are easily exhausted only a few and not all of the possible configurations can be achieved.
\item \textit{Configuration Invariability}  - The configuration of the network cannot be changed after the exhaustion of all the tokens or in other words it freezes after some amount of time.
\item \textit{Configuration Re-traceability} - We cannot revert back to a previous configuration after the arrival of the token.
\end{itemize}
To overcome some of these drawbacks we initially proposed a solution. The solution uses a new codebook design along with rules to generate node-shares and broadcast tokens from this codebook. It has been described in detail in the following chapter along with a mathematical proof and examples.

%% file: Selective.tex
\chapter{Selective Unlocking Mechanism}
\label{chap:Selectunlock}

We understood that one possible solution to the Token exhaustion problem would be to design tokens and node shares such that each token primarily unlocks only one key in one of the nodes in the network. We worked towards developing such a solution which would allow the distribution centre to control which key gets unlocked in which node in the network.
\section{Node share matrix design}
In order to realize the above mentioned solution, we propose to design a Node share matrix with the following distance property.\\
\begin{defn}
\textbf{Distance property}: Let $v$ be the number of keys. Then the hamming distance $H$ between any two pairs of the generated node share code words should lie between$[4,\:v\:-\:4]$. In other words, let $N\textsubscript{i}$ and $N\textsubscript{j}$ be two node share code words. Then,
\[
4\leq \: H(N\textsubscript{i},N\textsubscript{j}) \: \leq v\:-\:4
\]
\end{defn}
\subsection{Generation of Node share matrix with code words satisfying the Distance property}
We briefly present a simple method to generate node share code words satisfying the Distance property. For all future purposes we use the following notations:
\begin{itemize}
\item \textit{v} represents the number of keys
\item \textit{n} represents the maximum number of nodes that can be supported by a ‘\textit{v}’ key system
\item \textit{N} represents an $n \times v$ Node share matrix
\item $\textit{N}\textsubscript{i}$ represents the $i\textsuperscript{th}$ node share code word or the $i\textsuperscript{th}$ row of the Node share matrix N
\item \textit{S} is an $\textit{n} \times (\textit{v}/2)$  matrix such that \textit{N}= $[\textit{S}\;\;\textit{S}\textsuperscript{c}]$
\item $\textit{S}\textsubscript{i}$ represents the $i\textsuperscript{th}$ row in \textit{S}
\end{itemize}
We consider a $\textit{n} \times (\textit{v}/2)$ matrix \textit{S}.
\begin{enumerate}
\item To construct \textit{S} we start with a partial code word $\textit{S}\textsubscript{1}$ with equal number of zeros and ones i.e, $\textit{S}\textsubscript{1}$ has (\textit{v}/4) zeros and (\textit{v}/4) ones. $\textit{S}\textsubscript{1}$ is the first row of \textit{S}.
\item To obtain the other rows we permute the bits of $\textit{S}\textsubscript{1}$ and consider only the permutations which are not complementary to each other i.e, no two rows $\textit{S}\textsubscript{i}$ and $\textit{S}\textsubscript{j}$ should be complementary.
\item To obtain the Node share matrix, we concatenate \textit{S} and its complement.
\end{enumerate}
\underline{\textit{Note}}:
\begin{itemize}
\item It can be observed that the maximum number of such non complementary permutations equals $n$. Hence,
\[
n \:= \: (v/2)!/2(v/4)!(v/4)!
\]
\item The code words generated by this method are \textit{n} out of \textit{n} collusion secure.
\end{itemize}
\subsection{Proof that these code words satisfy the Distance property}
To prove the Distance property for these code words, we consider two partial code words $\textit{S}\textsubscript{i}$ and $\textit{S}\textsubscript{j}$. Without loss of generality, let us assume that $\textit{S}\textsubscript{j}$ is obtained by a permutation on the bits of $\textit{S}\textsubscript{i}$. Since $\textit{S}\textsubscript{i}$ and $\textit{S}\textsubscript{j}$ have equal number of 1s and 0s, the least hamming distance that could be obtained by a permutation is 2. Hence,
\[
H(S\textsubscript{i},S\textsubscript{j}) \: \geq 2
\]
Similarly, since $\textit{S}\textsubscript{i}$ and $\textit{S}\textsubscript{j}$ cannot be complements of each other, using the same equal number of 0s and 1s argument, we can prove that the maximum Hamming distance between them can be at most $(\textit{v}/2)\:-\:2$. Hence,
\[
H(S\textsubscript{i},S\textsubscript{j}) \: \leq (v/2) \:-\: 2
\]
Since $\textit{N}\textsubscript{i}$ and $\textit{N}\textsubscript{j}$ are just concatenations of $\textit{S}\textsubscript{i}$, $\textit{S}\textsubscript{i}\textsuperscript{c}$, $\textit{S}\textsubscript{j}$ and $\textit{S}\textsubscript{j}\textsuperscript{c}$ respectively, we have
\[
4\leq \: H(N\textsubscript{i},N\textsubscript{j}) \: \leq v\:-\:4
\]
Hence, the satisfaction of distance property has been proved for these code words.
\subsection{Illustration: Node share matrix generation for a 12 key system}
Since \textit{v}=12, we get \textit{n}=10 from the equation derived above. Let $\textit{S}\textsubscript{1}\;=\;[0\: 0\: 0\: 1\: 1\: 1]$, then \textit{S} is given by:\\
\[
\left({\begin{array}{cccccc}
0 & 0 & 0 & 1 & 1 & 1 \\
0 & 0 & 1 & 0 & 1 & 1 \\
0 & 0 & 1 & 1 & 0 & 1 \\ 
0 & 0 & 1 & 1 & 1 & 0 \\
0 & 1 & 0 & 0 & 1 & 1 \\
0 & 1 & 0 & 1 & 0 & 1 \\
0 & 1 & 0 & 1 & 1 & 0 \\ 
0 & 1 & 1 & 1 & 0 & 0 \\
0 & 1 & 1 & 0 & 1 & 0 \\
0 & 1 & 1 & 0 & 0 & 1 \\
\end{array} } \right)
\]
\\
The Node share matrix is then obtained as \textit{N} = $[\textit{S} \;\; \textit{S}\textsuperscript{c}]$ \\
\subsection{Node share distribution}
In this scheme, each node is given one node share. For example, the node share of node \textit{i} is generated using the $\textit{i}\textsuperscript{th}$ row of the Node share matrix i.e, $\textit{N}\textsubscript{i}$ becomes the node share code word for node \textit{i}.

\section{Token design}
Let us say that the centre wants to unlock key '\textit{j}' in node '\textit{i}' without unlocking any other key in any other node. Let the corresponding token be denoted by $\textit{T}\textsubscript{ij}$. Then $\textit{T}\textsubscript{ji}$ is given by:
\[
T\textsubscript{ji}\:=\:N\textsubscript{i} \oplus I\textsubscript{j}
\]
where $\textit{I}\textsubscript{j}$ is the $\textit{j}\textsuperscript{th}$ row of a $\textit{v} \times \textit{v}$ identity matrix \textit{I}. In other words, the $\textit{j}\textsuperscript{th}$ bit of $\textit{N}\textsubscript{i}$ is flipped to get $\textit{T}\textsubscript{ji}$.\\

\subsection{Proof of desired token behaviour}
In this proof, we use the rules for conditional visibility and invisibility of partitions which have been described earlier. Since only one bit of $\textit{N}\textsubscript{i}$ gets flipped to get $\textit{T}\textsubscript{ji}$, we have $H(\textit{N}\textsubscript{i}$,$\textit{T}\textsubscript{ji})\: =\: 1$. But we already know that,
\[
4\leq \: H(N\textsubscript{i},N\textsubscript{k}) \: \leq v\:-\:4 \;\;\forall \;k \:\neq\: i
\]
Since one bit flip to $\textit{N}\textsubscript{i}$ can at most reduce the Hamming distance between $\textit{N}\textsubscript{i}$ and $\textit{N}\textsubscript{k}$ by 1 or increase it by at most 1, we get
\[
3\leq \: H(T\textsubscript{ji},N\textsubscript{k}) \: \leq v\:-\:3 \;\;\forall \;k \:\neq\: i
\]
Since the Hamming distance between $\textit{T}\textsubscript{ji}$ and $\textit{N}\textsubscript{k}$ is at least 3, when $\textit{T}\textsubscript{ji}$ is stacked over $\textit{N}\textsubscript{k}$ there would be at least three columns satisfying the equation $[\:\bar{b} \:\; b\:]\textsuperscript{T}$.\\ \\
Similarly since the Hamming distance between $\textit{T}\textsubscript{ji}$ and $\textit{N}\textsubscript{k}$ is at most \textit{v} - 3, when  $\textit{T}\textsubscript{ji}$ is stacked  over $\textit{N}\textsubscript{k}$  there would be at least three columns satisfying the equation $[\:b \:\; b\:]\textsuperscript{T}$\\ \\
Hence there would be no unique equation that is revealed when $\textit{T}\textsubscript{ji}$ is stacked over $\textit{N}\textsubscript{k}$ $\forall k \neq i$ So no keys will be unlocked in any other node. However, when $\textit{T}\textsubscript{ji}$ is stacked over $\textit{N}\textsubscript{i}$, the column which is given by the equation $[\:\bar{b} \:\; b\:]\textsuperscript{T}$ becomes unique. Hence Key '\textit{j}' would be unlocked in node '\textit{i}'.\\ \\
\subsection{Illustration}
Suppose, we need to unlock key $5$ in node $6$ in a $12$ key system. We use the same Node share matrix presented before in this example. $T\textsubscript{56}$ equals:\\
\[T\textsubscript{56} = 
\left({\begin{array}{cccccccccccc}
0 & 1 & 0 & 1 & 1 & 1 & 1 & 0 & 1 & 0 & 1 & 0\\
\end{array} } \right)
\]
\\
By stacking $\textit{T}\textsubscript{56}$ over each $\textit{N}\textsubscript{i}$, it can be easily verified that Key $5$ will be unlocked only in node $6$ and no keys will be unlocked in any other node.\\
\subsection{Advantages of the proposed code book design}

\begin{itemize}

\item Many Tokens can be generated and it is possible to achieve any network configuration. In fact upto \textit{n}*\textit{v} different Tokens can be generated and $\textit{n}\textsuperscript{v}$ configurations can be achieved.

\item Node shares and Tokens are easy to generate.

\item Node shares are \textit{n} out of collusion secure i.e, even if all the node shares are stacked over each other, no partition would be revealed.

\item For larger values of \textit{v}, a much larger network can be supported.

\end{itemize}

%% file: Proposedscheme.tex
\chapter{Proposed scheme}
\label{chap:Proposedscheme}

The method of release of keys described in \cite{Karthik13} ensures dynamic reconfiguration. However, it poses the challenge of control over the node associations. In the previous chapter, we proposed a solution which incorporates control in forming these node associations. However, the code book design had other issues. The tokens generated are only 2 collusion secure which means that by combining more than two tokens, the eaves dropper could potentially learn the hidden partitions. Another major drawback of the previouly proposed solution, is the bandwidth limitations of the tokens. They do not support simultaneous release of multiple keys in different nodes and hence the tokens waste a lot of bandwidth. \\ \\
One of the common points that one could notice in the two schemes is that the partitions allotted to the keys are fixed and do not vary from one node share to another. The token design is constrained by the requirement to maintain and collusion security among tokens. The difficulty in incorporating controlled key release over this frame work stems primarily from static partition allotment which in turn forces any two tokens to be related. Our next proposal differs from these two schemes in this aspect. We vary the partitions to which different keys are allotted and design tokens which are unrelated to each other. We show that our solution achieves controlled and simultaneous key release without compromising the token security. 

\section{Description:}

The setup of the network is similar to the one described in Section 2. The network comprises of a center $D$ and nodes Node 1, Node 2, \dots, Node $n$ all within the transmission range of each other. 
Let $K\textsubscript{1}$, $K\textsubscript{2}$, $K\textsubscript{3}$,\dots, $K\textsubscript{v}$ be $v$ keys each of length $L\textsubscript{p}$, generated by the center. Let $n$ be the maximum number of nodes that can be supported by a '$v$'-key system and $M$ represent the number of preliminary shares given to each node, i.e, for any node $i$ , the set of node share code words would be $N\textsubscript{i}$ = { $N\textsubscript{i1}$, $N\textsubscript{i2}$ , \dots, $N\textsubscript{iM}$ }\\ \\
Each key $K\textsubscript{i}$ in the set of $v$ keys is divided into two halves $K\textsubscript{i1}$ and $K\textsubscript{i2}$. Each half is locked in a hidden partition. Hence the size of each codeword is $(1 \times 2v)$. The first half of the keys, $K\textsubscript{i1}\; \forall i\: = \:1, \:2, \:\dots,\:v$ are encoded using the first $v$ bits of the code word while the second half, $K\textsubscript{i2}\; \forall i\:=\:1,\:2,\:\dots,\:v$  are encoded using the next $v$ bits of the code word.

\section{Formulation}

Prior to the code word construction, each Key $K\textsubscript{i}$ is assigned a fixed number such that no two keys are assigned the same number or its complement in binary. It is implied that since each key is assigned a fixed number, both the halves of the key are implicitly assigned the same number. For example, in a $4$ key system, the following assignment could be made:
\begin{table}[h]
    \centering
    \caption{Key Table}
    \begin{tabular}{|c|c|}
    \hline
    {\bf Key} & {\bf Number Assigned} \\
    \hline
    K\textsubscript{1}  & 1               \\
    \hline
    K\textsubscript{2}  & 2               \\
    \hline
    K\textsubscript{3}  & 3               \\
    \hline
    K\textsubscript{4}  & 7               \\
    \hline
    \end{tabular}
    
\end{table}\\
$M$ is calculated as the number of bits required to represent the largest number in the key table. From the table shown above, it can be deduced that the value of $M$ is $3$ in this case.

\section{Code word representation}

Here we describe a convenient way to represent Node share code words. Consider a set of 3 preliminary share code words (which together are used to generate the node share for a particular node $i$) representing a 4 Key system (8 partitions numbered from 1 to 8). It must be noted that these set of code words have repetitive columns (Rule 1). Hence the node shares generated from these code words do not reveal any keys.
\begin{table}[h]
    \centering
    \begin{tabular}{cccccccc}

    1  & 0 & 0 & 0 & 0 & 0 & 0 & 1   \\
    1  & 1 & 1 & 0 & 1 & 1 & 0 & 1      \\
    1 & 0 & 1 & 1 & 0 & 1 & 1 & 1
    \end{tabular}
\end{table}\\
If these code words are read column wise, it is easy to see that they represent the following sequence of numbers:
\[
N\textsubscript{i} = 7 \; 2 \; 3 \; 1 \; 2 \; 3 \; 1 \; 7
\]
If it is assumed that the above stated Key table is used in the mapping, then this means that, in this set of code words the following mapping has taken place :
\begin{table}[h]
    \centering
    \caption{Partition Assignment}
    \begin{tabular}{|c|c|c|}
    \hline
    {\bf Key (Half)} & {\bf Assigned Number} & {\bf Partition No.}\\
    \hline
    K\textsubscript{11}  & 1 & 4              \\
    \hline
    K\textsubscript{12}  & 1 & 7               \\
    \hline
    K\textsubscript{21}  & 2 & 2              \\
    \hline
    K\textsubscript{22}  & 2 & 5               \\
    \hline
    K\textsubscript{31}  & 3 & 3              \\
    \hline
    K\textsubscript{32}  & 3 & 6              \\
    \hline
    K\textsubscript{41}  & 7 & 1               \\
    \hline
    K\textsubscript{42}  & 7 & 8               \\
    \hline
    \end{tabular}
    
\end{table}\\
A particular Key $K\textsubscript{j}$ in the node $i$, can be represented by the positions of its two halves. In the above example: \\

$K\textsubscript{1}\textsuperscript{i} = (4,7),\; K\textsubscript{2}\textsuperscript{i} = (2,7),\;K\textsubscript{3}\textsuperscript{i} = (3,6),\;K\textsubscript{4}\textsuperscript{i} = (1,8)$

\section{Unique mapping property}

If a particular key $K\textsubscript{i}$ in node $j$ is represented by the coordinates $(x,y)$  i.e,
\[
K\textsubscript{i}\textsuperscript{j}\;=\;(x\textsubscript{i}\textsuperscript{j},y\textsubscript{i}\textsuperscript{j})
\]
Then no other node has key $K\textsubscript{i}$ in the positions $(x\textsubscript{i}\textsuperscript{j},y\textsubscript{i}\textsuperscript{j})$ .This must be true for all $K\textsubscript{i}\textsuperscript{'}s$ . Hence,
\[
\{K\textsubscript{i}\textsuperscript{p} \neq  (x\textsubscript{i}\textsuperscript{j},y\textsubscript{i}\textsuperscript{j})\; \forall p \neq j\} \;and \;\forall\; i
\]
\section{Node share code word generation}
In order to generate node shares $N\textsubscript{i}$ satisfying the Unique mapping property, we define two matrices $C\textsubscript{1}$ and $C\textsubscript{2}$ which are both $(v \times v)$ matrices whose rows are used to construct the node share code words. Both matrices $C\textsubscript{1}$ and $C\textsubscript{2}$ are generated by the following algorithm. Let $C\textsubscript{i}[j]$ represent the $j\textsuperscript{th}$ row of the matrix $C\textsubscript{i}$.
\begin{algorithm}
\caption{Construction of C\textsubscript{1} and C\textsubscript{2}}
\begin{algorithmic}
\STATE Start with a $(1 \times v)$ random permutation ($R$) of the numbers in the key table 
\FOR{ $j \leftarrow 1 \; to \; v$}
\STATE $C\textsubscript{i}[j] \leftarrow R$
\STATE Circular right shift $R$ by $1$
\ENDFOR
\end{algorithmic}
\end{algorithm}
\begin{flushleft}
{\bf $Illustration :$} 
\end{flushleft}
Let the number of keys be 4. Refer to Table I for the mapping of keys. Let
$C\textsubscript{1}$ represent a matrix containing first half of node share code words formed using a random seed $R\textsubscript{1}$ say:
\[
R\textsubscript{1} =
\left[{\begin{array}{cccc}
7 & 1 & 2 & 3 
\end{array} } \right]
\]
$C\textsubscript{1}$ is then given by:
\[
C\textsubscript{1} =
\left[{\begin{array}{cccc}
7 & 1 & 2 & 3 \\
3 & 7 & 1 & 2 \\
2 & 3 & 7 & 1\\
1 & 2 & 3 & 7 
\end{array} } \right]
\]
Similarly, the other matrix $C\textsubscript{2}$ can be obtained with another random seed say 
\[
R\textsubscript{2} =
\left[{\begin{array}{cccc}
3 & 2 & 1 & 7 
\end{array} } \right]
\]
$C\textsubscript{2}$ is then given by:
\[
C\textsubscript{2} =
\left[{\begin{array}{cccc}
3 & 2 & 1 & 7 \\
7 & 3 & 2 & 1 \\
1 & 7 & 3 & 2 \\
2 & 1 & 7 & 3 
\end{array} } \right]
\]
After the obtaining both matrices $C\textsubscript{1}$ and $C\textsubscript{2}$, the node share code word for any node with ID $i$ is constructed as follows,
\[
Let \; q = Quotient(i/v) ; \hspace{2.0mm} r = (i\hspace{1.0mm} mod \hspace{1.0mm} v), \; then, \\
\]
\begin{equation}
N\textsubscript{i} = C\textsubscript{1}(q)\parallel C\textsubscript{2}(r)
\end{equation}
This procedure could generate a maximum of $v\textsuperscript{2}$ node share code words. Hence, at most $v\textsuperscript{2}$ nodes can be supported by a $v$ key system. It can be easily seen that the node shares generated by this procedure are $n$ out of $n$ collusion secure. Both random seeds $R\textsubscript{1}$ and $R\textsubscript{2}$ contain exactly one instance of every number in the key table. Thus every node share code word contains exactly two instances of every number in the key table. From Rule 1, one can infer that the code words have repetitive columns and they do not reveal any hidden partitions. \\ \\
\textit{Example:}\\ \\
Let us find the node share code word for Node 7. Since Node ID = 7 and $v\; =\; 4$, we have $q\: =\: 1$, $r \:=\: 3$. Then,
\[
N\textsubscript{7} =
\left[{\begin{array}{cccccccc}
3 & 7 & 1 & 2 & 2 & 7 & 1 & 3
\end{array} } \right]
\]
\textit{Note:} Appendix I proves that code words generated using this procedure satisfy the Unique Mapping property.

\section{Token design}
The unlocking of desired partitions in one or more nodes is done with the help of broadcast tokens. The tokens are fabricated in a way so as to unlock the corresponding partitions without the release of undesired partitions in nodes other than the target nodes. \\ \\
\textit{Illustration:} \\ \\
Referring to the example stated earlier, for a 4-key system let the node shares in two nodes i and j be given by: \\
\[
N\textsubscript{i} =
\left[{\begin{array}{cccccccc}
7 & 2 & 3 & 1 & 2 & 3 & 1 & 7
\end{array} } \right]
\]
\[
N\textsubscript{j} =
\left[{\begin{array}{cccccccc}
7 & 2 & 3 & 1 & 3 & 2 & 7 & 1
\end{array} } \right]
\]
Suppose we want to unlock the partitions corresponding to the number 7 ,i.e, we would like to release key $K\textsubscript{4}$  in nodes $i$ and $j$.  Then they are represented by the following points: 
\begin{equation}
K\textsubscript{4}\textsuperscript{i} = (1,8) \nonumber
\end{equation}
\begin{equation}
K\textsubscript{4}\textsuperscript{j} = (1,7) \nonumber
\end{equation}
This can be achieved effectively by the design of proper token. The token $T$ is designed as:
\[
T =
\left[{\begin{array}{cccccccc}
K\textsubscript{41} & K\textsubscript{R} & K\textsubscript{R} & K\textsubscript{R} & K\textsubscript{R} & K\textsubscript{R} & K\textsubscript{42}\textsuperscript{'} & K\textsubscript{42}\textsuperscript{'}
\end{array} } \right]
\]
\[
P =   
\left[{\begin{array}{cccccccc}
  P\textsubscript{1}\: &   P\textsubscript{2}\: &   P\textsubscript{3}\: & P\textsubscript{4}\: & P\textsubscript{5}\: & P\textsubscript{6}\: & P\textsubscript{7}\: & P\textsubscript{8}
\end{array} } \right]
\]
where $K\textsubscript{41}$ is the first half of $K\textsubscript{4}$ and $K\textsubscript{42}\textsuperscript{c}$ is the bit-compliment of the second half of $K\textsubscript{4}$, and $K\textsubscript{R}$ is a random $L\textsubscript{p}$ bit binary number.
$P$ represents the corresponding partitions. In this example, $K\textsubscript{41}$ is filled in partition 1 i.e $P\textsubscript{1}$. $K\textsubscript{42}\textsuperscript{c}$ is filled in partitions 7 and 8. The other partitions are filled with random $L\textsubscript{p}$ bit numbers $K\textsubscript{R}$\\ \\
When $T$ fuses with $N\textsubscript{i}$ and $N\textsubscript{j}$ , stack relation equations corresponding to $K\textsubscript{41}$ and $K\textsubscript{42}$ become unique. In this example, the equation for $K\textsubscript{41}$ will be $[\:b\; b\; b\; b\:]\textsuperscript{T}$ ( which corresponds to the number $[1\:1\:1\:1\:]\textsuperscript{T}$ ) whereas the stack equation for $K\textsubscript{42}$ is $[\:\bar{b}\; b\; b\; b\:]\textsuperscript{T}$ ( it corresponds to the number $[0\: 1\: 1\: 1]\textsuperscript{T}$ ) in both nodes $i$ and $j$. \\ \\
In other words, the stack equation of $P\textsubscript{1}$ in both nodes would be:
\[
\left[{\begin{array}{cccc}
b & b & b & b 
\end{array} } \right]\textsuperscript{T}
\]
and the stack equation of $P\textsubscript{7}$  in node $j$ and the stack equation of $P\textsubscript{8}$ in node $i$ would be:
\[
\left[{\begin{array}{cccc}
\bar{b} & b & b & b 
\end{array} } \right]\textsuperscript{T}
\]
In all the other partitions (other than $P\textsubscript{1}$ and $P\textsubscript{7}$ in node $j$ and $P\textsubscript{1}$ and $P\textsubscript{8}$ in node $i$), no key would be released. This is because the token holds a random number in these partitions which is not related to any of the keys residing in them. Hence the stack equations would be invalid for these partitions.\\ \\
Further, since the unique mapping property is satisfied, no other node will correspond to these points $(1,8)$ and $(1,7)$ in the Key $K\textsubscript{4}$ space . In all the other nodes, the stack equations would be invalid for at least one of the halves of the key $K\textsubscript{4}.$ Hence at least one of the halves cannot be unlocked in any other node. (In fact both halves cannot be unlocked in every other node. It will be subsequently proved). This results in the simultaneous controlled release of Key $K\textsubscript{4}$ only in nodes i and j.
\[
K\textsubscript{4}\;=\;K\textsubscript{41}\;\parallel\;K\textsubscript{42}
\]
\section{General procedure}
\begin{itemize}

\item Suppose $K\textsubscript{i}$ is to be unlocked in nodes $p$ and $q$. Let
$K\textsubscript{i}\textsuperscript{p} = (x\textsubscript{i}\textsuperscript{p},y\textsubscript{i}\textsuperscript{p})\;and\; K\textsubscript{i}\textsuperscript{q} = (x\textsubscript{i}\textsuperscript{q},y\textsubscript{i}\textsuperscript{q})$

\item Then if at least one of the coordinates are equal i.e,  $(x\textsubscript{i}\textsuperscript{p}\: =\: x\textsubscript{i}\textsuperscript{q}$  or $y\textsubscript{i}\textsuperscript{p}\: =\: y\textsubscript{i}\textsuperscript{q}$, the token is constructed as follows :

$K\textsubscript{i1}$ is filled in partition numbers $x\textsubscript{i}\textsuperscript{p}$ and $x\textsubscript{i}\textsuperscript{q}$, $K\textsubscript{i2}\textsuperscript{c}$ is filled in partition numbers $y\textsubscript{i}\textsuperscript{p}$ and $y\textsubscript{i}\textsuperscript{q}$.

\item This idea can easily be extended to more than two nodes as well, provided at least one of the coordinates is the same for all nodes in the privileged set. In other words, suppose if the key $K\textsubscript{i}$ is to be released in the set of nodes $S\: =\:{(x\textsubscript{1},y\textsubscript{1}),\: (x\textsubscript{2},y\textsubscript{2}),\: \dots,\: (x\textsubscript{N},y\textsubscript{N}) }$ such that $x\textsubscript{1}\:=\: x\textsubscript{2}\:=\: \dots x\textsubscript{N}\: =\: x$ and $y\textsubscript{i}$ are all distinct, a single token $T$ can be constructed as follows:\\

$K\textsubscript{i1}$ is filled in partition number $x$ and  $K\textsubscript{i2}\textsuperscript{c}$ is filled in partition numbers $y\textsubscript{j}\; \forall\;  j\:=\:1,\:2,\: ...,\:N$. A similar approach can be used for the alternative scenario where all the $y$ coordinates are the same and all the $x$ coordinates are distinct for the members of the privileged set.\\

\textit{Note:} Appendix II proves that the above stated procedure indeed achieves controlled and simultaneous key release.
\end{itemize}
 
\begin{itemize}

\item It must be noted in the above stated general procedure that if both $(x\textsubscript{i}\textsuperscript{p}$ $\neq$ $x\textsubscript{i}\textsuperscript{q}$   and $y\textsubscript{i}\textsuperscript{p}$ $\neq$ $y\textsubscript{i}\textsuperscript{q}$, then the key will be unlocked in the desired nodes p and q as well as two other nodes corresponding to the points $(x\textsubscript{i}\textsuperscript{p},y\textsubscript{i}\textsuperscript{q} )$ and $(x\textsubscript{i}\textsuperscript{q},y\textsubscript{i}\textsuperscript{p} )$.  Hence two separate messages must be sent to avoid this. 

\item In other words, a single message can unlock a key in a set of nodes or points $(x\textsubscript{1},y\textsubscript{1})$, $(x\textsubscript{2},y\textsubscript{2})$, \dots, $(x\textsubscript{N},y\textsubscript{N})$ if and only if one of the coordinates ($x\textsubscript{i}$ or $y\textsubscript{i}$) are equal for all the points i.e the points lie on a line parallel to the x or y axis in a 2-D Grid. If this is not the case, then the set of points can be broken into disjoint subsets each satisfying the above stated requirement, and a token can be constructed for each subset to unlock the key in all the nodes of that subset. The number of tokens required will then be equal to the number of subsets. In the worst case, the number of subsets would be equal to the number of points N which would in turn translate to N tokens.

\item Further, if one of the coordinate say $\textbf{x}$ is fixed, then $\textbf{y}$ can take at most $\textbf{v}$ values. This means that a single token can unlock one key in a maximum of $v$ nodes simultaneously.

\item It must also be noted that two different keys $K\textsubscript{i}$ and $K\textsubscript{j}$ can also be unlocked simultaneously in two different nodes $p$ and $q$ respectively, provided both the coordinates are such that $x\textsubscript{i}\textsuperscript{p}$ $\neq$ $x\textsubscript{j}\textsuperscript{q}$   and $y\textsubscript{i}\textsuperscript{p}$ $\neq$ $y\textsubscript{j}\textsuperscript{q}$. This can be easily extended to more than two nodes as well.

\item Another observation that can be made is that more than one key can be unlocked simultaneously in a particular node with a single token. In fact up to all the $\textbf{v}$ keys can be unlocked in a particular node with a single token.

\end{itemize}

%% file: Tokenconstruction.tex
\chapter{Token construction}
\label{chap:Tokenconstruction}

One of the goals of the reconfiguration protocol is to minimize the number of tokens and achieve the desired configuration. In the previous section, we showed that a single token can unlock the same key or different keys simultaneously in many nodes. This property of the design can be used to reduce the number of tokens. This section describes a sub optimal greedy algorithm which makes maximum utilization of each token. The algorithm uses two functions which are analysed first.

\section{Choosing points from a Key grid}

Any desired configuration can be achieved by unlocking different subsets of keys in different nodes. Since each key is represented by a point in that particular key space/grid depending on the node at which it is to be unlocked, the whole configuration can be considered as a set of $v$ key grids $G\textsubscript{1}$, $G\textsubscript{2}$ $\dots$, $G\textsubscript{v}$ with each key grid $G\textsubscript{i}$ containing a set of points where each point represents a node at which the Key K\textsubscript{i} has to be unlocked. Let the desired configuration be represented by $G$, then, 
\[G = \{G\textsubscript{1},\; G\textsubscript{2},\; \dots\; ,G\textsubscript{v}\}  \] \\
where, $G\textsubscript{i}\;=\;\{p\textsubscript{1},\; p\textsubscript{2},\; \dots\;,\;p\textsubscript{n\textsubscript{i}}\}\; and\; |G\textsubscript{i}|\;=\;n\textsubscript{i} $\\
Let $p\textsubscript{i}[x]$ and $p\textsubscript{i}[y]$ represent the $x$ and $y$ coordinates of any point $p\textsubscript{i}$. Let $S\textsubscript{X}[x\textsubscript{k}]$ be the set of points from a given key grid with their x coordinate equal to $x\textsubscript{k}$. Similarly, let $S\textsubscript{Y}[y\textsubscript{k}]$ be the set of points from a given key grid with their y coordinate equal to $y\textsubscript{k}$.
\begin{defn}
A valid subset of points in a key grid $G\textsubscript{i}$ is any subset of points belonging to $G\textsubscript{i}$ with all points in the subset having the same x coordinate or the same y coordinate. In other words, all points in a valid subset lie on lines parallel to either the x axis or the y axis. \\
\end{defn}
\begin{algorithm}
\caption{$find\_largest\_valid\_subset(G\textsubscript{i})$}
\begin{algorithmic}
\STATE $S\textsubscript{X} \leftarrow \emptyset$
\STATE $S\textsubscript{Y} \leftarrow \emptyset$
\FOR{$j \leftarrow 1$ to $ |G\textsubscript{i}| $}
\STATE $P\textsubscript{j} = G\textsubscript{i}(j)$
\STATE $S\textsubscript{X}(P\textsubscript{j}[x])\;=\;S\textsubscript{X}(P\textsubscript{j}[x])\bigcup P\textsubscript{j}$
\STATE $S\textsubscript{Y}(P\textsubscript{j}[y])\;=\;S\textsubscript{Y}(P\textsubscript{j}[y])\bigcup P\textsubscript{j}$
\ENDFOR
\STATE $S\textsubscript{Xbest}\;=\;S\textsubscript{X}(\lambda)\;\;s.t\;\;\lambda\;=\; \argmin\limits_{x} |S\textsubscript{X}(x)|$
\STATE $S\textsubscript{Ybest}\;=\;S\textsubscript{Y}(\gamma)\;\;s.t\;\;\gamma\;=\; \argmin\limits_{y} |S\textsubscript{Y}(y)|$
\IF{$|S\textsubscript{Xbest}| > |S\textsubscript{Ybest}|$}
\STATE $return\;\;S\textsubscript{Xbest}$
\ELSE
\STATE $return\;\;S\textsubscript{Ybest}$
\ENDIF
\end{algorithmic}
\end{algorithm}
Algorithm 2 finds the largest valid subset in any key grid $G\textsubscript{i}$. It has a running time of $O(|G\textsubscript{i}|)$ which is $O(|G|/v)$ on an average.

\section{Building a token}
The following algorithm (Algorithm 3) fills the partitions in a token with key halves. It implements the design principles described in the previous section. Let the $T$ be the token and let $T[i]$ represent the $i\textsuperscript{th}$ partition in $T$ i.e, $P\textsubscript{i}$. The algorithm takes in an argument a set $S$ which is a valid subset in the key space of a particular key say $K\textsubscript{i}$. 
\begin{algorithm}
\caption{$Build\_token(T,S,K\textsubscript{i})$}
\begin{algorithmic}
\FOR {$i \leftarrow 1$ to $|S|$}
\STATE $p\textsubscript{i} \leftarrow S[i]$
\STATE $T(p\textsubscript{i}[x]) \leftarrow K\textsubscript{i1}$
\STATE $T(p\textsubscript{i}[y]) \leftarrow K\textsubscript{i2}\textsuperscript{c}$
\ENDFOR
\end{algorithmic}
\end{algorithm}\\
The running time of the algorithm is $O(|S|)$ which is reasonably small.

\section{Token construction}
The algorithm described in this subsection (Algorithm 4) uses the Algorithms 2 and 3 to construct tokens and eventually achieve the desired configuration $G$. Before presenting the pseudo code, two important definitions which are used by the algorithm are stated below.
\begin{defn}
Consider a key grid containing a set of points. A grid span $\beta\textsubscript{R,G\textsubscript{i}}$ of any set of points $R$, is defined as \\ 
\[\beta\textsubscript{R,G\textsubscript{i}}\:=\:\{p \in G\textsubscript{i} \; s.t \; \exists c \in R \;\; p[x]=c[x]\;\:OR\:\;p[y]=c[y]\}\]
In general, for any two sets $A$ and $B$,
\[\beta\textsubscript{B,A}\:=\:\{p \in A \; s.t \; \exists c \in B \;\; p[x]=c[x]\;\:OR\:\;p[y]=c[y]\}\]
\end{defn}
\begin{defn}
The operation $\ast$ is defined on two sets $A$ and $B$ as follows, \\ 
\[A \ast B = A\;-\;\beta\textsubscript{B,A}\]
In other words, the operation $\ast$ finds the set of points in $A$ which do not lie in the grid span of $B$.
\end{defn}

The algorithm progresses from one key grid to the next and at each step it finds the largest valid subset among points in the key grid which do not belong to the grid span of any of the previously computed largest valid subsets of the other key grids. The algorithm assumes that the desired configuration $G$ is available. At each iteration, a set $R$ stores points (pairs of partitions) which are to be filled by keys in order to construct the Token. The set $R$ is updated from one key grid $G\textsubscript{i}$ to the next by adding all points belonging to the largest valid subset of $G\textsubscript{i+1}\ast R$ i.e, by combining the largest valid subset computed from points in $G\textsubscript{i+1}$ which do not lie in the grid span of any of the previously computed largest valid subsets. \\ \\
\begin{algorithm}
\caption{Token\_construction}
\begin{algorithmic}
\STATE $j \leftarrow 1$
\WHILE{$G \neq \emptyset$}
\STATE $T\textsubscript{j} \leftarrow Random\_Init(|T|)$
\STATE $R \leftarrow \emptyset$
\FOR {$i \leftarrow 1$ to $v$}
\IF{$No\:further\: points\: can\: be\: added\: to\: T$} 
\STATE $Transmit(T\textsubscript{j})$  
\STATE Jump to START
\ELSE
\IF{$G\textsubscript{i} \neq \emptyset$}
\STATE $R\textsubscript{i}\leftarrow find\_largest\_valid\_subset(G\textsubscript{i}\ast R)$
\STATE $G\textsubscript{i}\leftarrow G\textsubscript{i}\;-\;R\textsubscript{i}$
\STATE $Build\_token(T\textsubscript{j},R\textsubscript{i},K\textsubscript{i})$
\STATE $R\leftarrow R\bigcup R\textsubscript{i}$
\ENDIF
\ENDIF
\STATE $i\leftarrow i\;+\;1$
\ENDFOR
\STATE $Transmit(T\textsubscript{j})$ 
\STATE START : $j\leftarrow j\;+\;1$
\ENDWHILE
\end{algorithmic}
\end{algorithm}\\
The algorithm has a run time complexity of $O(|G|\textsuperscript{2})$. 
Our model assumes a powerful distribution centre which constructs tokens. Hence the burden of computation falls on the distribution centre and computational costs at the nodes are minimal.

%% file: Security.tex
\chapter{Security}\label{chap: Security}
\section{Collusion resistance}
Any set of nodes which do not belong to the privileged set cannot collaborate to reconstruct key halves. Let us consider a scenario where a token releases key $K\textsubscript{i}$ in a privileged set. Some bits of the key halves Ki1 and Ki2 would be released in the other nodes. Can they collude and reconstruct the key halves ? \\ \\
The answer is that it would be difficult to do so in polynomial time.  Each token is randomly initialised  and hence the random numbers filled will be differ from one partition to another within the same token. So the possible "bit values" of $K\textsubscript{i1}$ and $K\textsubscript{i2}$ that could be revealed (in a node which does not belong to the privileged set of $K\textsubscript{i}$ ) would also vary from one node to another. The relative position of these bits would also vary because the key halves would occupy different partitions in different nodes. Since both the bit values as well as their positions differ from one node to another, it would be difficult for the nodes to collude because they would not know which bit values would have to be combined to get the key halves. Due to variation of the bit positions, they should not be able to guess the order of bit combination as well. Further there is an added complexity because only some bits of $K\textsubscript{i1}$ and $K\textsubscript{i2}$ will be released and they would be revealed along with a union of some random bits. These random bits would also vary from one node to another because of the random token initialisation. So two colluding nodes cannot find out which bits in the released Union (some bits of $K\textsubscript{i1}$ or $K\textsubscript{i2}$  + some random bits) belong to the key halves because both ( bits of $K\textsubscript{i1}$ or  $K\textsubscript{i2}$  and random bits) would vary from one node to another.

\section{Token security}
Since the tokens consist of two partitions for each key-node pair which are randomly distributed, the formation of any unique columns (if any) by stacking more than one token together will not lead to the release of a key. When a key is released in a particular node i.e, when the partitions corresponding to the point represented by the node are filled with the key halves, no other subsequent token would have the same key halves in the same two partitions. Hence unique columns cannot be formed by stacking because the partitions that a particular key occupies keeps varying from one token to another. The tokens formed are unrelated to each other and thus they are collusion-secure. The scheme can be made less sensitive to physical node capture by introducing some additional steps. Since both halves of a key are unlocked separately to obtain the key, we can ensure that the node deletes all bits of the node share that lie in the union of these two partitions. The node then stores the union of the two partitions. Before processing tokens, the node simply removes all bits in the token that lie in the union of the two partitions. The other partitions remain intact and the stack relations do not change. The attacker who captures the node can only recover the unlocked key and the union of its corresponding two partitions. If he is able to identify the hidden partitions, then he could use the previously transmitted tokens to obtain information about other keys which could have resided in the compromised partitions. By storing only the union of the two hidden partitions, we ensure that the attacker cannot obtain the two halves separately in polynomial time or gain information about other keys from the past or future token transmissions.

%% file: Discussion.tex
\chapter{Discussions}
\label{chap:Discussions}


\section{Properties of Tokens}
Each token can carry at most $v$ keys which are not necessarily unique. For each token, we define the total instance of a particular Key $K\textsubscript{i}$ as the number of nodes in which the token releases the key $K\textsubscript{i}$. It is represented by the symbol $I\textsubscript{K\textsubscript{i}}$. If the key $K\textsubscript{i}$ is not released by the token in any node, $I\textsubscript{K\textsubscript{i}}$ is set to zero. The tokens can be classified based on their bandwidth utilization, into two types  : Efficient and Inefficient. The classification depends on the total number of instances of all keys released by the token i.e, the classification is done using the quantity $|K|\;=\;\sum\textsubscript{i}\;I\textsubscript{K\textsubscript{i}}$\\ \\
{\bf Efficient Tokens:} Tokens which release atleast $v$ instances of one or more keys are termed as Efficient ($|K|\: \geq \: v$). A single token can simultaneously release at most $2v-2$ instances in total of two or more keys i.e $|K|\textsubscript{max}\:=\:2v-2$. For instance, simultaneous release of $2v-2$ instances of two keys (say $K\textsubscript{i}$ and $K\textsubscript{j}$) is possible if two non overlapping valid subsets (of $K\textsubscript{i}$ and $K\textsubscript{j}$ respectively) each of size $v-1$ ($I\textsubscript{K\textsubscript{i}}\:=\:v-1$ and $I\textsubscript{K\textsubscript{j}}\:=\:v-1$) are grouped together into one token. As an example, in a 3 key system, suppose the privileged set of $K\textsubscript{i}$ is $S\textsubscript{i}$ = $\{(1,4),(1,5)\}$ and the  privileged set of $K\textsubscript{j}$ is $S\textsubscript{j}$ = $\{(2,6),(3,6)\}$, then the simultaneous release of $K\textsubscript{i}$ and $K\textsubscript{j}$ in $S\textsubscript{i}$ and $S\textsubscript{j}$ respectively, can be achieved by a single token. This is because the privileged sets are non overlapping i.e the partitions occupied by these privileged sets do not intersect. In this example, although the total size of the transmitted token is only three key lengths (since $v\:=\:3$), it is able to unlock $|K|\:=\:4$ instances of keys (2 instances of each key). Such tokens where $|K|\: \geq \: v$ are Efficient because by transmitting only $vL\textsubscript{p}$ bits, the token is able to release useful information amounting to $|K|L\textsubscript{p}$ bits which is greater than or equal to the amount of bits transmitted. \\ \\
{\bf Inefficient Tokens:} Tokens which release less than $v$ instances of one or more keys are termed Inefficient ($|K|\: \leq \: v$). The reasons for their inefficiency are obvious from the previous discussion on Efficient Tokens. Inefficient tokens waste bandwidth because most partitions are filled with random numbers which only ensure security and exclusive key release. If a token is heavily underutilized, then transmitting the keys through other key distribution mechanisms could save bandwidth. The process of assigning keys to different groups plays an important role in minimizing the number of Inefficient tokens. For an arbitrary configuration with a large number of nodes, the number of inefficient tokens is generally small. Usually, the final few tokens that are transmitted to complete a given configuration, are inefficient.
\section{Comparisons with Tree based Broadcast encryption}
Tree based broadcast encryption schemes impose a hierarchical structure on the network and inherently form groups by doing so. Efficient low memory tree based schemes require each node to store $O(logn)$ keys. If the controller decides to form $N\textsubscript{G}$ multicast groups in the network, the group key establishment phase would require the transmission of only $O(N\textsubscript{G})$ messages provided these groups are the same as the inherently formed ones. However, if arbitrary multicast groups are required, $O(m)$ messages would have to be transmitted for each group of average size $m$. This implies that formation of $N\textsubscript{G}$ arbitrary groups would require the transmission of $O(mN\textsubscript{G})$ messages. On the other hand, with the proposed scheme, any arbitrary multicast group with a size $m$ greater than $v$ would necessarily have overlapping subsets. Hence, in general less than $m$ messages would be required to establish the group key. Moreover, the multicast groups can grow simultaneously as well. Assuming each token adds $k$ nodes to each group, one token would release keys in $kN\textsubscript{G}$ nodes on an average. The total number of messages required would be $O(m/k)$ which is much lesser when compared to that of the Tree based Broadcast encryption schemes. However, for a network of size $v\textsuperscript{2}$, each node would have to store $O(v)$ keys which is higher than the storage requirement of low memory broadcast encryption methods.

\section{Forward secrecy}
Nodes which leave a multicast group should not have acess to future conversations. This can be achieved by creating one additional key not known to the leaving member of the group. Fast group key revocation with minimum number of messages requires the group members to form valid subsets in multiple key grids thereby imposing additional constraints on the design. Furthermore, if the design is changed to include such partially overlapping subsets to form one specific multicast group, it could reduce the flexibility in forming other groups linked to the same key.
Consider a scenario of a network with four nodes 1, 2, 3 and 4. Now let keys $K\textsubscript{1}$, $K\textsubscript{2}$ and $K\textsubscript{3}$ be released such that the following multicast group associations are formed :
\begin{table}[h]
    \centering
    \caption{Initial group associations}
    \begin{tabular}{|c|c|}
    \hline
    {\bf Group} & {\bf Group key} \\
    \hline
    G\textsubscript{1} = \{1,2,3\}  & K\textsubscript{1}\\
    \hline
    G\textsubscript{2} = \{1,3,4\}  & K\textsubscript{2}\\
    \hline
    G\textsubscript{3} = \{2,3,4\}  & K\textsubscript{3}\\
    \hline
    \end{tabular}
    
\end{table}
Now assume node 1 leaves the network. This would imply that node 1 should no longer have the priviledge to future conversations within groups $G\textsubscript{1}$ and $G\textsubscript{2}$. This means that nodes 2 and 3  should share a unique common multicast key while the same is true for nodes 3 and 4. Individual nodes could form all possible functional combinations of keys released in them. This would imply the mapping shown in Table.~\ref{Table: Initial node key associations}.
\begin{table}[h]
    \centering
    \caption{Initial Node - key associations}\label{Table: Initial node key associations}
    \begin{tabular}{|c|c|}
    \hline
    {\bf Node} & {\bf Revealed Keys} \\
    \hline
    1 & K\textsubscript{1}, K\textsubscript{2}, H(K\textsubscript{1}, K\textsubscript{2})\\
    \hline
    2 & K\textsubscript{1}, K\textsubscript{3}, H(K\textsubscript{1}, K\textsubscript{3})\\
    \hline
    3 & K\textsubscript{1}, K\textsubscript{2}, K\textsubscript{3}, All possible hashes\\
    \hline
    4 & K\textsubscript{2}, K\textsubscript{3}, H(K\textsubscript{2}, K\textsubscript{3})\\
    \hline
    \end{tabular}

\end{table}\\
Since node 1 leaves the network all keys unlocked in 1 cannot be used further. The corresponding mapping is shown in Table.~\ref{Table: Associations after Node 1 leaves}.
\begin{table}[h]
    \centering
    \caption{Node - key associations after revocation of node 1}\label{Table: Associations after Node 1 leaves}
    \begin{tabular}{|c|c|}
    \hline
    {\bf Node} & {\bf Revealed Keys} \\
    \hline
   2 & K\textsubscript{3}, H(K\textsubscript{1}, K\textsubscript{3})\\
    \hline
   3 & K\textsubscript{2},K\textsubscript{3}, H(K\textsubscript{2},K\textsubscript{3}), H(K\textsubscript{1},K\textsubscript{3}), H(K\textsubscript{1},K\textsubscript{2},K\textsubscript{3})\\
    \hline
    4 & K\textsubscript{3}, H(K\textsubscript{2},K\textsubscript{3}) \\
    \hline
    \end{tabular}
    
\end{table}\\
The new group associations are given by $G\textsubscript{1}\textsuperscript{`}\: =\: \{2,3\}$, $G\textsubscript{2}\textsuperscript{`} \:=\: \{3,4\}$, $G\textsubscript{3}\textsuperscript{`}\: =\: \{2,3,4\}$ with group keys H($K\textsubscript{1},K\textsubscript{3}$), H($K\textsubscript{2},K\textsubscript{3}$) and $K\textsubscript{3}$ respectively. Node 1 has been removed from Groups $G\textsubscript{1}$ and $G\textsubscript{2}$ without the release of any new keys. The other members of each group still continue to be a part of their respective groups. \\ \\
However, if node 3 decides to leave all its groups, then all the keys in the other three nodes would have to be updated which is in stark contrast to the previous scenario where no new keys were released. In some situations it is also possible to use even the compromised keys to create group keys for the modified groups without revealing any information about the new group keys to the leaving node. \\ \\
Another possible solution is to form all possible multicast group associations with a unique key for each association.This would mean that the keys corresponding to subsets which contain the leaving node can be discarded and other keys can be used. This pre-designed approach would impose constraints on flexibility of group associations in the network and increase the number of pre defined distinct keys $K\textsubscript{i}$. \\ \\
We can define depth as the total number of nodes which have left the network or their groups. This would imply that depth can vary from 1 to $n-1$ for multicast group associations. Future work should revolve around minimizing the number of disctinct keys, messages and support forward secrecy at different depth values. 

%% file: Simulations.tex
\chapter{Simulations}
\label{chap:Simulations}
We analyse the performance of the scheme based on the number of tokens that have to be transmitted to configure the network. We study a scenario where the system contains a fixed number of nodes and we find the number of tokens required to form all possible multicast groups of different sizes. It must be noted here that the sizes of the networks simulated are small and as a result, the number of required tokens should tend to be small as well. Consequently, significant bandwidth gains cannot be expected because the final few tokens that are transmitted, often tend to be inefficient. However, in larger networks, most of the transmitted tokens are likely to be Efficient and wastage of bandwidth over all tokens is less likely. A seven key, five node network is used in both scenarios. Since the network has 5 nodes, there are ten groups of size 2, ten of size 3 and five groups of size 4. The five node share code words are generated according to the procedure described in section 3.5. Further, it is assumed that every node computes and stores some function or hash of all possible combinations of keys that were unlocked in that node. \\ \\
{ \bf Group size 2:} By unlocking keys according to the following table, a unique group key can be established for every possible group of size 2. The algorithm described in section 4 was used to construct tokens and we found that all ten groups, each of size 2, can be created by the transmission of four broadcast tokens.
\begin{table}[h]
    \centering
    \caption{Multicast groups of size 2}
    \begin{tabular}{|c|c|c|c|}
    \hline
    {\bf Group} & {\bf Group key} & {\bf Group} & {\bf Group key}\\
    \hline
    \{1,2\}  & H(K\textsubscript{1},K\textsubscript{2}) & \{2,4\} & H(K\textsubscript{1},K\textsubscript{6}) \\
    \hline
    \{1,3\}  & H(K\textsubscript{1},K\textsubscript{3}) & \{2,5\} & H(K\textsubscript{2},K\textsubscript{5}) \\
    \hline
    \{1,4\}  & H(K\textsubscript{1},K\textsubscript{4}) & \{3,4\} & H(K\textsubscript{1},K\textsubscript{7}) \\
    \hline
    \{1,5\}  & H(K\textsubscript{2},K\textsubscript{4}) & \{3,5\} & H(K\textsubscript{5},K\textsubscript{7}) \\
    \hline
    \{2,3\}  & H(K\textsubscript{1},K\textsubscript{5}) & \{4,5\} & H(K\textsubscript{4},K\textsubscript{7}) \\
    \hline
    \end{tabular}
    
\end{table} \\
{\bf Group size 3:} For every group of size 3, a unique group key can be obtained as a function of two primary keys $K\textsubscript{i}$ and $K\textsubscript{j}$ for some $i$ and $j$. Table 7 shows the key mappings for every possible triad of nodes. Similar to the previous case, we found that all the ten triads can be created by transmitting 4 tokens. \\ \\
\begin{table}[h]
    \centering
    \caption{Multicast groups of size 3}
    \begin{tabular}{|c|c|c|c|}
    \hline
    {\bf Group} & {\bf Group key} & {\bf Group} & {\bf Group key}\\
    \hline
    \{1,2,3\}  & H(K\textsubscript{1},K\textsubscript{2}) & \{1,4,5\} & H(K\textsubscript{3},K\textsubscript{4}) \\
    \hline
    \{1,2,4\}  & H(K\textsubscript{1},K\textsubscript{3}) & \{2,3,4\} & H(K\textsubscript{1},K\textsubscript{5}) \\
    \hline
    \{1,2,5\}  & H(K\textsubscript{2},K\textsubscript{3}) & \{2,3,5\} & H(K\textsubscript{2},K\textsubscript{5}) \\
    \hline
    \{1,3,4\}  & H(K\textsubscript{1},K\textsubscript{4}) & \{2,4,5\} & H(K\textsubscript{3},K\textsubscript{5}) \\
    \hline
    \{1,3,5\}  & H(K\textsubscript{2},K\textsubscript{4}) & \{3,4,5\} & H(K\textsubscript{4},K\textsubscript{5}) \\
    \hline
    \end{tabular}
    
\end{table}\\
{\bf Group size 4:} The key-group mappings shown in Table 8, when implemented yield five groups after the broadcast of four tokens (Similar to the previous two cases). \\ \\
\begin{table}[h]
    \centering
    \caption{Multicast groups of size 4}
    \begin{tabular}{|c|c|c|c|}
    \hline
    {\bf Group} & {\bf Group key} \\
    \hline
    \{1,2,3,4\}  & K\textsubscript{1} \\
    \hline
    \{1,2,3,5\}  & K\textsubscript{2} \\
    \hline
    \{1,2,4,5\}  & K\textsubscript{3} \\
    \hline
    \{1,3,4,5\}  & K\textsubscript{4} \\
    \hline
    \{2,3,4,5\}  & K\textsubscript{5} \\
    \hline
    \end{tabular}
    
\end{table}\\
The simulations reveal that only 4 tokens were required to form as large as 10 groups signalling improvement over traditional static key establlishment schemes and broadcast encryption schemes.

%% file: Conclusion.tex
\chapter{Conclusion}
\label{chap:Conclusion}
In this thesis report we have presented a design methodology for controlling the release of protected encryption keys in different nodes, to enforce a dynamic re-configuration of the virtual wireless network, with the help of broadcast tokens. With the proposed node share and token design, any arbitrary virtual multicast configuration can be realized through a sequence of carefully designed tokens. Both the node shares and the broadcast tokens are collusion resistant. Further extensions have been discussed, to incorporate forward secrecy, which comes with a price of compromising on the network flexibility. A paper on our proposal has been submitted to the ACM WiSEC 2014 conference in manchester, UK and it is pending approval.

%% file: Appendix.tex
\begin{appendices}
\renewcommand{\thechapter}{\Roman{chapter}}
\chapter{Unique mapping property}
The following proof demonstrates that when the general node share code generation procedure described in section 3.5 is followed, the resulting node share codewords satisfy the the Unique mapping property. Consider any two node shares which are distributed to two nodes $i$ and $j$. They are characterised by the tuples $(q\textsubscript{1}, r\textsubscript{1})$ and $(q\textsubscript{2}, r\textsubscript{2})$ respectively. It must be noted that the two tuples cannot be equal because of the fact that they represent two distinct nodes. Hence, at least one of the parameters, either $q$ or $r$ must be different. \\ \\
{\bf Case 1:} Assuming that the parameter $r$ is different but $q$ remains the same between the two tuples, then the corresponding code word halves are $C\textsubscript{2}(r\textsubscript{1})$ and $C\textsubscript{2}(r\textsubscript{2})$ respectively. But $C\textsubscript{2}(r\textsubscript{2})$ is obtained from $C\textsubscript{2}(r\textsubscript{1})$ by a certain number of circular shifts less than or equal to $v-1$ ( $v$ circular shifts will result in the same sequence again). Since all the assigned key numbers (in the code word representation) are distinct in both $C\textsubscript{2}(r\textsubscript{1})$ and $C\textsubscript{2}(r\textsubscript{2})$, no column will have the same number as both entries when $C\textsubscript{2}(r\textsubscript{1})$ and $C\textsubscript{2}(r\textsubscript{2})$ are stacked one below the other. Hence the $y$ coordinate of every key (partition of the second half of every key) will differ in the Nodes $i$ and $j$. \\ \\
{\bf Case 2:} The parameter $q$ is different but $r$ remains the same between the two tuples. The argument is similar to the one presented in Case 1 \\ \\
{\bf Case 3:} Both $q$ and $r$ are different, follows trivially from Case 1 and Case 2

\chapter{Simultaneous key release}
Let $S\;=\; \{(x\textsubscript{1},y\textsubscript{1}),\:(x\textsubscript{2},y\textsubscript{2}), \dots, \:(x\textsubscript{N},y\textsubscript{N}\} $ be a $valid \;subset$ of nodes in which a key say $K\textsubscript{i}$ has to be released. Then $S$ is defined as the privileged set of key $K\textsubscript{i}$. The following proof demonstrates that the Token design procedure described in section 3.7 indeed unlocks the key $K\textsubscript{i}$ in every node belonging to the privileged set $S$. Further, the proof also shows that any node which does not belong to the privileged set cannot obtain any information about the key $K\textsubscript{i}$. Let $T$ be a token that is generated using the token design procedure described in section 3.7 to unlock $K\textsubscript{i}$ exclusively in the privileged set $S$.

\section{\bf At a node belonging to the privileged set} 
Let the node $j$ belong to the privileged set. Let it be represented by the point $(x,y)$ with respect to key $K\textsubscript{i}$. Since the token was randomly initialised, every other partition apart from partitions $x$ and $y$ contain random numbers. No other key (apart from $K\textsubscript{1}$) could be unlocked at this node because of invalid stack relations. Recall that in the node share $N\textsubscript{j}$ $K\textsubscript{i1}$ is filled in the partition $x$ and $K\textsubscript{i2}$ is filled in the partition $y$. Since both $K\textsubscript{i1}$ and $K\textsubscript{i2}$ were assigned the same key table number, they share the same stack equation in $N\textsubscript{i}$. However, one of the partitions (either $x$ or $y$) in the token $T$ is filled with the complement of the corresponding key half. Fusion of the token with $N\textsubscript{j}$ results in unique stack equations for $K\textsubscript{i1}$ and  $K\textsubscript{i2}$. Hence both $K\textsubscript{i1}$ and $K\textsubscript{i2}$ can be extracted and concatenated to obtain key $K\textsubscript{i}$ 

\section{\bf At a node which does not belong to the privileged set} 
Consider a node $A$ which does not belong to the privileged set. Let it be represented by the point $(x\textsubscript{a},y\textsubscript{a})$ with respect to key $K\textsubscript{i}$. Let us assume that the $x$ coordinate be the same for all nodes in the privileged set $S$ i.e, $S\;=\; \{(x\textsubscript{k},y\textsubscript{1}),\:(x\textsubscript{k},y\textsubscript{2}), \dots, \:(x\textsubscript{k},y\textsubscript{N}\} $ where the $x$ coordinate equals $x\textsubscript{k}$. \\ \\
{\bf Case 1 : $x\textsubscript{a}\; = \; x\textsubscript{k}$ } \\
In this case the $y$ coordinate of $A$ must be different from all the $y\textsubscript{j}$s of the privileged set $S$. This is because if $y\textsubscript{a} = y\textsuperscript{j}$ for some $y\textsubscript{j}$ in the privileged set $S$, then the point $(x\textsubscript{a},y\textsubscript{a})$ will lie inside the privileged set S which is not possible. \\
Hence, $y\textsubscript{a} \;\neq \;y\textsubscript{j} \;\forall \; j\;=\;1,2,\dots, N$ \\
If the token design procedure is followed, then the partition corresponding to $y\textsubscript{a}$ will be filled with a random number which is not equal to $K\textsubscript{i2}\textsuperscript{c}$.  Let this random number be $K\textsubscript{R}$.  Since $K\textsubscript{R} \;\neq \; K\textsubscript{i2}\textsuperscript{c}$, some bits of $K\textsubscript{R}$ must be the same as those of $K\textsubscript{i2}\textsuperscript{c}$. Let these positions in the partition be represented by the set $B$. 
Since the partition corresponding to $y\textsubscript{a}$ is filled with a random number $K\textsubscript{R}$ , the stack equations would become invalid and the second half $K\textsubscript{i2}$ cannot be unlocked in node $A$. When node $A$ tries to unlock $K\textsubscript{i1}$, the bits corresponding to the partition $x\textsubscript{a}$ i.e $P\textsubscript{X\textsubscript{a}}$ are released along with the bits present in the positions denoted by the set $B$. In other words when the node $A$ tries to unlock $K\textsubscript{i1}$, it instead gets $\{P\textsubscript{X\textsubscript{a}} \bigcup B\}$ i.e, a union of both these sets. It cannot determine the partition $P\textsubscript{X\textsubscript{a}}$ from this union. Hence $K\textsubscript{i1}$ would not be unlocked.
Thus both halves $K\textsubscript{i1}$ and $K\textsubscript{i2}$ cannot be unlocked. No other key would be unlocked as well because every other partition would have invalid stack relations. \\ \\
{\bf Case 2 : $x\textsubscript{a}\; \neq \;x\textsubscript{k}\; and\; y\textsubscript{a}\;=\;y\textsubscript{j}\;for\;some\; y\textsubscript{j}\; in\; S$ }\\ 
Using similar arguments as in Case 1, it can be clearly seen that the partition corresponding to $x\textsubscript{a}$ i.e $P\textsubscript{X\textsubscript{a}}$ will be filled with some random number $K\textsubscript{R1}$ . This will ensure that $K\textsubscript{i1}$ cannot be unlocked by node $A$. Since $K\textsubscript{R1}\;\neq\;K\textsubscript{i1} $ some bits in $K\textsubscript{R1}$ will be equal to those in $K\textsubscript{i1}\textsuperscript{c}$. These will again form a non-null set $B$ which will ensure that $K\textsubscript{i2}$ is not unlocked. (Similar to Case 1). No other key would be unlocked as well because every other partition would have invalid stack relations. \\ \\
{\bf Case 3 : $x\textsubscript{a}\;\neq\; x\textsubscript{k}\;and\;y\textsubscript{a}\;\neq\;y\textsubscript{j}\;\forall \;j\;=\;1,2,\dots,\;N$} \\
In this case, both the partitions corresponding to $x\textsubscript{a}$ and $y\textsubscript{a}$ will be filled with random numbers. This yields invalid stack equations and both halves cannot be unlocked by node $A$. No other key would be unlocked as well because every other partition would have invalid stack relations. \\ \\
No other cases are possible. This proves that the key $K\textsubscript{i}$ is not unlocked in any other node which does not belong to the privileged set. Similar arguments can be given when for the scenario where all the $y$ coordinates of the privileged set $S$ are equal and all the $x$ coordinates are distinct. 

\end{appendices}